\def\thalf{{\textstyle{3\over2}}}

\def\hlf{{{1\over2}}}
\def\thlf{{{3\over2}}}

\def\>{\rangle}
\def\<{\langle}

\def\beq{\begin{equation}}
\def\eeq{\end{equation}}

\def\beqy{\begin{eqnarray}}
\def\eeqy{\end{eqnarray}}
\def\l{\lambda} 
\def\al{\alpha_\lambda} 
\def\ar{\alpha_\rho} 
\def\bl{\beta_\lambda} 
\def\br{\beta_\rho} 
\documentclass[aps,showpacs,floatfix]{revtex4}
\usepackage{graphicx,epsfig,rawfonts,pictex,color,multirow,longtable}
\newlength{\dinwidth}
\newlength{\dinmargin}
\begin{document}
\thispagestyle{empty}
\title{Hyperfine Mixing and the Semileptonic Decays of Double-Heavy Baryons in a Quark Model}
\author{W. Roberts$^{1}$ and Muslema Pervin$^{2}$}
\affiliation{$^1$ Department of Physics, Florida State University, Tallahassee, FL 32306\\
$^2$ Physics Division,~Argonne National Laboratory, Argonne, IL-60439}

\begin{abstract}
The semileptonic decays of the lowest-lying double-heavy baryons is treated in a quark model. For the $\Xi_{bb}$, hyperfine mixing in the spin wave function leaves the total rate for decay into the lowest lying daughter baryons essentially unchanged, but changes the relative rates into the $\Xi_{bc}$ and $\Xi_{bc}^\prime$. The same pattern is obtained in the decays of the $\Omega_{bb}$. For the $\Xi_{bc}$, this mixing leads to factor of about 17 suppression in the decay rate to the $\Xi_{cc}^*$, and a factor of two suppression in the total decay rate. For the $\Omega_{bc}$, the decay to the $\Omega_{cc}^*$ is suppressed by a factor of more than 30 from the unmixed case, and the total decay rate is decreased to about 40\% of the decay rate obtained when mixing is ignored.
\end{abstract}
\pacs{12.39.-x, 12.39.Jh, 12.39.Pn, 13.30.-a, 13.30.Ce, 14.20.-c}
\maketitle
\setcounter{page}{1}

%%%%%%%%%%%%%%%%%%%%%%%%%%%%%%%%%%%%%%%%%%%%%%%%%
\section{Introduction and Motivation}

\label{sec:intro}

In a recent manuscript \cite{pr}, we examined the spectra of heavy baryons,
including those containing more than one heavy quark, in a quark model. Among
the baryons containing two different flavors of heavy quarks, one aspect of
particular interest was the possibility of hyperfine mixing among the baryons
made of scalar (spin zero) and axial vector (spin one) heavy diquarks. The
result from that model was that the mixing among the baryons was quite large, and the authors went on to 
suggest that this mixing would have significant impact on other properties of
these states. In the present manuscript, we examine the effects of this hyperfine
mixing on the semileptonic decays of baryons containing two heavy quarks.

The semileptonic decays of doubly-heavy baryons have been treated
by a number of authors. White and Savage \cite{savage} examined
these decays using heavy quark spin symmetry, while Sanchis-Lozano \cite{sanchis} applied the heavy quark effective theory to a discussion of both baryons and mesons containing two heavy quarks. Albertus and collaborators
\cite{albertus}, Flynn and Nieves \cite{flynn}, Ebert {\it et al.} \cite{ebert},
Hernandez, Nieves and Verde-Velasco \cite{hernandez} and
Faessler {\it et al.} \cite{faessler} have examined
these decays in quark models of different kinds. Guo {\it et al.} \cite{guo} obtain estimates for the decay widths using a
Bethe-Salpeter formalism. In the present work, we use a quark model that has
been applied to the semileptonic decays of baryons containing a single heavy
quark \cite{prca,prcb}.

In the ground states of the doubly-heavy baryons $\Xi_{QQ}$ and $\Omega_{QQ}$, where $Q=b$ or $Q=c$, the pair of heavy quarks is expected to form a diquark with spin one. In the limit of infinitely massive heavy quarks, the spin $s_h$ of this heavy diquark is expected to decouple from that of the light quark, leading to the so-called heavy quark spin symmetry (HQSS). One implication is that there should be a pair of nearly-degenerate states composed of the diquark with spin 1 and the light quark with spin $\hlf$. These two states are denoted $\Xi_{QQ}\,\, (\Omega_{QQ})$, having total angular momentum $\hlf$, and $\Xi_{QQ}^* \,\, (\Omega_{QQ}^*)$ with total angular momentum $\thlf$. For the states $\Xi_{bc}$ and $\Omega_{bc}$, because the heavy quarks are not identical, in addition to the diquark with spin 1, the diquark in these states can also have total spin zero, leading to another state with total angular momentum $\hlf$. This state is usually denoted $\Xi_{bc}^\prime\,\,(\Omega_{bc}^\prime)$ to distinguish it from the state containing the symmetric heavy diquark. These states and their properties are summarized in Table \ref{states}. 
\begin{center}
\begin{table}\label{states}
\caption{The lowest lying states among the baryons containing two heavy quarks. $s_h$ is the total spin of the heavy quarks. The spin wave functions in column four assume that hyperfine mixing can be ignored. When this mixing is not ignored, the spin wave functions of column five result. All masses are in GeV.} 
\begin{tabular}{|c|c|c|c|c|c|}\hline
State(Mass) & Unmixed Flavor & $s_h$& \multicolumn{2}{c|}{Spin Wave Function} & $J^P$  \\\cline{4-5}
 & Wave Function & (Unmixed) & Unmixed & Mixed & \\\hline
$\Xi_{cc}(3.676)$& $ccq$ & 1 &$\chi^\lambda$& N/A & $\hlf^+$ \\\hline
$\Xi_{cc}^*(3.753)$& $ccq$ & 1 & $\chi^S$ & N/A & $\thlf^+$ \\\hline
$\Xi_{bb}(10.340)$& $bbq$ & 1 & $\chi^\lambda$& N/A &$\hlf^+$ \\\hline
$\Xi_{bc}(7.011)$ & $\frac{1}{\sqrt{2}}(cb+bc)q$ & 1 & $\chi^\lambda$ &$0.92\chi^\lambda+0.39\chi^\rho$&
$\hlf^+$ \\\hline
$\Xi_{bc}^*(7.074)$& $\frac{1}{\sqrt{2}}(cb+bc)q$ & 1 & $\chi^S$ & $\chi^S$&$\thlf^+$ \\\hline
$\Xi_{bc}^\prime(7.047)$& $\frac{1}{\sqrt{2}}(cb-bc)q$ & 0 & $\chi^\rho$ & $0.92 \chi^\rho-0.39 \chi^\lambda$&$\hlf^+$ \\\hline

$\Omega_{cc}(3.815)$ & $ccs$ &1& $\chi^\lambda$& N/A &$\hlf^+$\\\hline
$\Omega_{cc}^*(3.876)$ & $ccs$ &1& $\chi^S$ & N/A &$\thlf^+$\\\hline
$\Omega_{bb}(10.454)$ & $bbs$ &1& $\chi^\lambda$& N/A &$\hlf^+$\\\hline
$\Omega_{bc}(7.136)$ & $\frac{1}{\sqrt{2}}(cb+bc)s$ &1& $\chi^\lambda$
&$0.91 \chi^\lambda+0.41\chi^\rho$&$\hlf^+$\\\hline
$\Omega_{bc}^*(7.187)$ & $\frac{1}{\sqrt{2}}(cb+bc)s$ &1& $\chi^S$& $\chi^S$& $\thlf^+$\\\hline
$\Omega_{bc}^\prime(7.165)$ & $\frac{1}{\sqrt{2}}(cb-bc)q$ &0& $\chi^\rho$ & $0.91\chi^\rho-0.41\chi^\lambda$& $\hlf^+$\\\hline

\end{tabular}
\end{table}

\end{center}

The $\Xi_{bc}$ and $\Omega_{bc}$ wave functions in column four of Table \ref{states} are obtained when the hyperfine interaction that induces mixing between heavy diquarks with different spins is suppressed. When this mixing is turned on, the spin wave functions of column 5 result in the model described in \cite{pr}. Note that the question of hyperfine mixing only applies to the states comprised of three different quark flavors. HQSS arguments suggest that this hyperfine mixing should be small, but the model calculation of \cite{pr} yields the large mixings shown in column five.

\section{The Model}

\subsection{Baryon Wave Functions}

Our discussion of the semileptonic decays of a baryon in our quark model begins with a description of the way in which a baryon wave function is constructed. In our model, a baryon wave function is described in terms of a totally antisymmetric color wave function, multiplying flavor, space and spin wave functions. We use $\phi$ to denote flavor wave functions, $\chi$ for spin, $\psi$ for space, and $\Psi$ for both the spin-space and spin-space-flavor wave functions. The spin-space wave function written for each state is partially determined by its flavor wave function. For flavor wave functions that are (anti)symmetric under exchange of the first two quarks, the spin-space wave function must also be (anti)symmetric under exchange of the first two quarks.

The total spin of the three spin-$\hlf$ quarks in the baryon can be either $\thlf$ or $\hlf$.  The
spin wave functions for the maximally stretched state in each case are 
\begin{eqnarray}
\chi_{3/2}^S(+3/2)  &=& |\uparrow\uparrow\uparrow\rangle, \nonumber\\
\chi_{1/2}^\rho(+1/2)  &=& 
\frac{1}{\sqrt{2}}(|\uparrow\downarrow\uparrow\rangle - 
|\downarrow\uparrow\uparrow\rangle), \nonumber\\
\chi_{1/2}^\lambda(+1/2)  &=&- 
\frac{1}{\sqrt{6}}(|\uparrow\downarrow\uparrow
\rangle  + |\downarrow\uparrow\uparrow\rangle -2|
\uparrow\uparrow\downarrow\rangle), \nonumber
\end{eqnarray}
where $S$ labels the state as totally symmetric, while $\lambda/\rho$ denotes
the mixed symmetric states that are symmetric/antisymmetric under the exchange
of quarks $1$ and $2$.

The Jacobi coordinates are chosen to be
\beq
{\bf \rho}= \frac{1}{\sqrt2}({\bf r}_1 - {\bf r}_2)
\eeq
and 
\beq
{\bf \lambda} =\sqrt{\frac{2}{3}}\left(\frac{m_1{\bf r}_1 + m_2{\bf r}_2}{m_1+m_2} - 
{\bf r}_3\right).
\eeq
${\bf \rho}$ is proportional to the separation of quarks 1 and 2, while ${\bf \l}$ is proportional to the separation between the third quark and the center of mass of the \{12\} pair of quarks.

Since we are interested in only the lowest-lying states with $J^P=\hlf^+$ or $\thlf^+$, the spatial wave function is  written as
\begin{equation}
\Psi({\bf \rho}, {\bf \lambda}) = \psi({\bf \rho}) \psi({\bf \lambda}),
\end{equation}
with
\beq\label{hoa}
\psi({\bf r})= \left[\frac{\alpha}{\sqrt{\pi}}\right]^\thalf e^{-\frac{\alpha^2r^2}{2}}.
\eeq
This form is an approximation of the form used in \cite{pr}, and amounts to truncating the expansion basis used in that work after the first term in the expansion. For the states that we are treating in this manuscript, this is a very good approximation. We note also that this spatial wave function is fully symmetric under permutation of any pair of quarks, a property that we will exploit shortly. With this form, the wave function for a state with $J^P=\hlf^+$ is written
\beq\label{wfn1}
\Psi^{\hlf^+}=\Psi({\bf \rho}, {\bf \lambda})\left(a\chi^\rho+b\chi^\lambda\right),
\eeq
while that for a state with $J^P=\thlf^+$ is
\beq\label{wfn3}
\Psi^{\thlf^+}=\Psi({\bf \rho}, {\bf \lambda})\chi^S.
\eeq

\subsection{Transition Matrix Elements}

We denote the currents responsible for the semileptonic decays we're interested in as $\Gamma$, and focus the discussion on the spin matrix elements, as these determine, to a large extent, the structure of the form factors for transitions between ground state baryons. Since we treat these matrix elements by assuming that the first two quarks are spectators in the decay, we have
\beq
\langle\chi^\lambda|\Gamma|\chi^\rho\rangle=\langle\chi^\rho|\Gamma|\chi^\lambda\rangle=0.
\eeq
For form factors describing decays between baryons that are antisymmetric in the first two quarks, the matrix elements
\beq
\langle\chi^\rho|\Gamma|\chi^\rho\rangle\equiv\Gamma_{\rho\rho}
\eeq
are needed. These $\Gamma_{\rho\rho}$ matrix elements are the ones used in \cite{prca} for the calculation of the form factors describing the semileptonic decays of heavy $\Lambda_Q$ baryons. For form factors describing decays between baryons that are symmetric in the first two quarks, the matrix elements
\beqy
\langle\chi^\lambda|\Gamma|\chi^\lambda\rangle&\equiv&\Gamma_{\lambda\lambda},\nonumber\\
\langle\chi^S|\Gamma|\chi^\lambda\rangle&\equiv&\Gamma_{S\lambda},\nonumber\\
\eeqy
are needed, and have been used to obtain the form factors for heavy $\Omega_Q$ decays in \cite{prcb}.

The equations for the form factors for the semileptonic transitions of a baryon with $J^P=\hlf^+$ to a daughter baryon with $J^P=\hlf^+$ are given in Appendix B.1.a of \cite{prcb} for a daughter baryon whose flavor wave function is symmetric in the first two quarks, and in Appendix C.1.a of \cite{prca} for a daughter baryon whose flavor  is antisymmetric in the first two quarks. For a daughter baryon with $J^P=\thlf^+$, the form factors are shown in Appendix B.8.a of \cite{prcb}. These form factors were obtained assuming that, for a quark transition $Q\to Q^\prime$ in the semileptonic decay $A\to B$, quark $Q$ is the third quark in the parent baryon $A$, and quark $Q^\prime$ is the third quark in the daughter baryon $B$. 

For the transitions that we consider here, the active quark is usually not the third quark in the wave functions that we construct. However, by performing a permutation on the full wave function, the active quark can be made to be the third quark. When such a permutation is carried out, the spatial wave function may be rewritten as
\beq
\Psi({\bf \rho}, {\bf \lambda})= \Psi({\bf \rho}^\prime, {\bf\lambda}^\prime),
\eeq
where $\rho^\prime$ and $\lambda^\prime$ are defined in the permuted basis. For example, for the $\{312\}$ ordering of quarks (the `default' ordering is $\{123\}$), the Jacobi coordinates are 
\beqy
{\bf \rho}^\prime&=& \frac{1}{\sqrt2}({\bf r}_3 - {\bf r}_1),\nonumber\\
{\bf \lambda}^\prime &=&\sqrt{\frac{2}{3}}\left(\frac{m_3{\bf r}_3 + m_1{\bf r}_1}{m_3+m_1} - 
{\bf r}_2\right).
\eeqy
This redefinition of the Jacobi coordinates is accompanied by a `rescaling' of the wave function size parameters, $\al$ and $\ar$. The new size parameters are 
\beq
\bl=\sqrt{B^2\ar^2+D^2\al^2},\,\,\,\br=\ar\al/\bl.
\eeq
$B$ and $D$ are determined by the permutation required in the wave function. For the $\{312\}$ ordering of quarks, for instance, $B=\sqrt{3}/2$ and $D=m_2/(m_1+m_2)$. The spatial wave function written in terms of $\rho,\,\,\lambda,\,\,\ar$ and $\al$ has unit overlap with the wave function written in terms of $\rho^\prime,\,\,\lambda^\prime,\,\,\br$ and $\bl$.

The spin wave functions are also modified by the permutation of the quarks. Of course, the fully symmetric wave function, $\chi^S$, remains invariant under any permutation. The other two representations mix with each other, with the mixing coefficients determined by the particular permutation. For the $\{312\}$ ordering, the results are
\beqy
\chi^\rho&=&-\frac{1}{2}\left(\chi^{\rho^\prime}-\sqrt{3}\chi^{\lambda^\prime}\right),\nonumber\\
\chi^\lambda&=&-\frac{1}{2}\left(\sqrt{3}\chi^{\rho^\prime}+\chi^{\lambda^\prime}\right),
\eeqy
where the primes indicate that the wave functions are symmetric or antisymmetric under exchange of the quark labelled `3' with the quark labelled `1' in the original wave function.

\subsection{$bbq\to cbq$}

For the parent baryon with flavor wave function $bbq$, with $q=u, \,\,d$ or $s$, the space-spin wave function takes the form of Eq. (\ref{wfn1}) with $a=0$ and $b=1$, for $J=\hlf$. When the quarks are rearranged so that the flavor wave function is $qbb$, the spin wave function becomes $$-\frac{1}{2}\left(\sqrt{3}\chi^{\rho^\prime}+\chi^{\lambda^\prime}\right)$$
where the prime superscript indicates that the wave functions are symmetric or antisymmetric under exchange of the first two quarks in the new wave function, namely $q$ and $b$ in this case. 

The flavor wave function for the daughter baryon $cbq$ may be written in one of two ways $cbq$ or $qcb$, with the spin-space wave function written to match. In the $cbq$ form, the spin wave functions are usually taken to be
\beq
|cbq,\hlf^+\rangle=\chi^\lambda,\,\,\,\,|cbq,\hlf^{+\prime}\rangle=\chi^\rho,\,\,\,\,|cbq,\thlf^+\rangle=\chi^S.
\eeq
If hyperfine mixing is allowed, the spin wave function of a mixed spin one-half state is given by Eq. (\ref{wfn1}), with $a$ and $b$ determined by the size of the mixing matrix elements. To calculate the form factors, the quarks are rearranged into the order $qbc$, and the spin part of the wave function of Eq. (\ref{wfn1}) becomes
\beq
\Psi^{\hlf^+}=\frac{1}{2}\left[\left(a-\sqrt{3}b\right)\chi^{\rho^\prime}-\left(\sqrt{3} a+b\right)\chi^{\lambda^\prime}\right].
\eeq

The transitions between the ground state $bbq$ and the lowest lying spin-$\hlf$ $cbq$ states therefore require the matrix elements
\beq
\langle cbq,\hlf^+|\Gamma|bbq,\hlf^+\rangle=-\frac{\sqrt{3}}{4}\left(a-\sqrt{3}b\right)\Gamma_{\rho\rho}+\frac{1}{4}\left(\sqrt{3}a+b\right)
\Gamma_{\lambda\lambda}.
\eeq
Similarly,
\beq
\langle cbq,\thlf^+|\Gamma|bbq,\hlf^+\rangle=-\frac{1}{2}\Gamma_{S\lambda}.
\eeq
Note that this latter matrix element is the same for any representation chosen for the $|cbq,\thlf^+\rangle$.
If mixing between the possible configurations is ignored, the spin wave function of the daughter baryon has either $a=1$ or $b=1$, with the latter corresponding to the lowest lying state. In this case, the transition to this state requires the matrix element
\beq\label{mea}
\langle cbq,\hlf^+|\Gamma|bbq,\hlf^+\rangle=\frac{3}{4}\Gamma_{\rho\rho}+
\frac{1}{4}\Gamma_{\lambda\lambda},
\eeq
while that to the second $\hlf^+$ state ($a=1,\,\,b=0$) requires the matrix elements
\beq\label{meb}
\langle cbq,\hlf^{+\prime}|\Gamma|bbq,\hlf^+\rangle=\frac{\sqrt{3}}{4}\left(\Gamma_{\lambda\lambda} -\Gamma_{\rho\rho}\vphantom{\frac{3}{4}}\right).
\eeq

If the daughter baryon is written as $qcb$, but with the spin wave function still written as $a\chi^\rho+b\chi^\lambda$, then the $\hlf^+$ state has spin wave function $\chi^\rho$ ($a=1$ in Eq. (\ref{qcb})), and the $\hlf^{+\prime}$ state has spin wave function $\chi^\lambda$ ($b=1$ in Eq. (\ref{qcb})). In this case, the wave function of the parent baryon must be permuted as discussed above, while that of the daughter must be permuted to $qbc$. The transition matrix elements required in this case are 
\beqy \label{qcb}
\langle qcb,\hlf^+|\Gamma|bbq,\hlf^+\rangle&=&-\frac{\sqrt{3}}{4}\left(a+\sqrt{3}b\right)\Gamma_{\rho\rho}-\frac{1}{4}\left(\sqrt{3}a-b\right)
\Gamma_{\lambda\lambda},\nonumber\\
\langle qcb,\thlf^+|\Gamma|bbq,\hlf^+\rangle&=&-\frac{1}{2}\Gamma_{S\lambda}.
\eeqy

\subsection{$cbq\to ccq$}

Much of the discussion of the previous subsection can be applied to the semileptonic decays of the baryons with flavor content $cbq$. We assume that the spin wave function of the parent baryon can be written as in Eq. (\ref{wfn1}). After the appropriate rearrangements in both the parent and daughter baryon, the required matrix elements are 
\beqy \label{cbq}
\langle ccq,\hlf^+|\Gamma|cbq,\hlf^+\rangle&=&\frac{\sqrt{3}}{4}\left(a+\sqrt{3}b\right)\Gamma_{\rho\rho}-\frac{1}{4}\left(\sqrt{3}a-b\right)
\Gamma_{\lambda\lambda},\nonumber\\
\langle ccq,\thlf^+|\Gamma|cbq,\hlf^+\rangle&=&\frac{1}{2}\left(\sqrt{3}a-b\right)\Gamma_{S\lambda}.
\eeqy

\subsection{Effects of Mixing}

The matrix elements shown in Eqs. (\ref{mea}) and (\ref{meb}) assume that there is negligible mixing between the $\Xi_{bc}$ and $\Xi_{bc}^\prime$ (or $\Omega_{bc}$ and $\Omega_{bc}^\prime$) states, and that these states are expressed in the $cbq$ basis. If these states are written in the $qcb$ basis, the work of \cite{pr} indicates that the lowest-lying $J^P=\hlf^+$ state has a spin wave function that is predominantly $\chi^\rho$ (antisymmetric in the $q$ and $c$ quarks), with a small admixture of $\chi^\lambda$. In terms of the wave functions of the $cbq$ basis [Eq. (\ref{wfn1})], this corresponds to $a\approx 1/2$ and $b\approx\sqrt{3}/2$. In the limit that the mixing between the two states expressed in the $qcb$ basis is suppressed, `$\approx$' become `$=$',  and the transition matrix elements then become
\beqy
\langle qcb,\hlf^+|\Gamma|bbq,\hlf^+\rangle&=&\frac{\sqrt{3}}{4}\left(\Gamma_{\lambda\lambda}-\Gamma_{\rho\rho}\right),\nonumber\\
\langle qcb,\hlf^{+\prime}|\Gamma|bbq,\hlf^+\rangle&=&-\frac{1}{4}\left(3\Gamma_{\rho\rho}-\Gamma_{\lambda\lambda}\right),\nonumber\\
\langle qcb,\thlf^+|\Gamma|bbq,\hlf^+\rangle&=&-\frac{1}{2}\Gamma_{S\lambda},\nonumber\\
\langle ccq,\hlf^+|\Gamma|qcb,\hlf^+\rangle&=&-\frac{\sqrt{3}}{2}\Gamma_{\rho\rho},\nonumber\\
\langle ccq,\thlf^+|\Gamma|qcb,\hlf^+\rangle&=&0.
\eeqy
In the unmixed $qcb$ basis, the $|qcb,\hlf^{+\prime}\rangle$ has spin wave function $\chi^\lambda$, and the relevant matrix elements are
\beqy
\langle ccq,\hlf^+|\Gamma|qcb,\hlf^{+\prime}\rangle&=&-\frac{1}{2}\Gamma_{\lambda\lambda},\nonumber\\
\langle ccq,\thlf^+|\Gamma|qcb,\hlf^{+\prime}\rangle&=&\Gamma_{S\lambda}.
\eeqy
However, it is unlikely that these decays will ever be observed, as the electromagnetic decays $| qcb,\hlf^{+\prime}\rangle \to |qcb,\hlf^+\rangle+\gamma$ will dominate the spectroscopy of these states. The mixing therefore has a large impact on the semileptonic decays of the lowest lying $qcb$ states, leading to significant suppression of the possible decay modes.

\section{Numerical Results and Conclusion}

Table \ref{sldresults} shows the results that we obtain for the semileptonic decay rates of the lowest lying baryons containing two heavy quarks. In this table, columns 1 and 5 identify the decay and columns 2 and 6 show the decay rate obtained if hyperfine mixing in the $cbq$ states is neglected. Columns 3 and 7 show the results when these states are written as $qcb$, but with mixing still ignored, while columns 4 and 8 show the results obtained when hyperfine mixing is allowed, in either basis. For decays of the $\Xi_{bb}$ and $\Omega_{bb}$ to the spin-$\thlf$ states, the decay rate is independent of the representation ($cbq$ or $qcb$) of the daughter baryon.
\begin{center}
\begin{table}\label{sldresults}
\caption{Results for semileptonic decays in units of $10^{10}s^{-1}$. $cbq$ indicates that the spin wave function of the $\Xi_{bc}$ or $\Omega_{bc}$ states are taken to be symmetric or antisymmetric in the $b$ and $c$ quarks, while $qcb$ indicates that they are symmetric in the $q$ and $c$ quarks. Unmixed indicates that hyperfine mixing is ignored in the wave function.}
\begin{tabular}{|c|c|c|c|c|c|c|c|}\hline
Decay & $cbq$ unmixed & $qcb$ unmixed& $cbq$ mixed & Decay & $cbq$ unmixed & $qcb$ unmixed& $cbq$ mixed \\\hline
$\Xi_{bb}\to\Xi_{bc}\ell\nu$ & 0.72 & 0.19 & 0.29 & $\Omega_{bb}\to\Omega_{bc}\ell\nu$ & 0.88 & 0.24 & 0.33 \\\hline
$\Xi_{bb}\to\Xi_{bc}^\prime\ell\nu$ & 0.47 & 1.00 & 0.90 & $\Omega_{bb}\to\Omega_{bc}^\prime\ell\nu$ & 0.56 & 1.20 & 1.11 \\\hline
$\Xi_{bb}\to\Xi_{bc}^*\ell\nu$ & 4.64 & 4.64 & N/A & $\Omega_{bb}\to\Omega_{bc}^*\ell\nu$ & 4.66 & 4.66 & N/A \\\hline
$\Xi_{bc}\to\Xi_{cc}\ell\nu$ & 1.54 & 2.26 & 2.18 & $\Omega_{bc}\to\Omega_{cc}\ell\nu$ & 1.36 & 2.01 & 1.95 \\\hline
$\Xi_{bc}\to\Xi_{cc}^*\ell\nu$ & 3.56 & 0.00 & 0.20 & $\Omega_{bc}\to\Omega_{cc}^*\ell\nu$ & 3.87 & 0.00 & 0.12 \\\hline

\end{tabular}

\end{table}
\end{center}

The table shows that the total decay rates of the $\Xi_{bb}$ and $\Omega_{bb}$ states are independent of the representation chosen for the states. These total decay rates are also independent of mixing, but the relative decay rates into the two possible spin-$\hlf$ daughter baryons is significantly affected by mixing.

For the $\Xi_{bc}$ and $\Omega_{bc}$ mixing has a very significant effect on the total decay rates, as well as on the decay rates to exclusive channels. For the $\Xi_{bc}$, for instance, the total decay rate obtained when mixing is ignored (in the $cbq$ basis) is $5.10\times 10^{10} s^{-1}$. When the basis is changed to $qcb$, this rate decreases to slightly less than one half of this value ($2.26\times 10^{10} s^{-1}$). When mixing is included, the rate increases slightly from this latter value to $2.38\times 10^{10} s^{-1}$. 

The effect of mixing on the rate $\Xi_{bc}\to\Xi_{cc}$ is significant, but the effect on the rate to $\Xi_{cc}^*$ is very dramatic. Without mixing, in the $cbq$ basis, this decay channel dominates the semileptonic decays of the $\Xi_{bc}$, accounting for about 70\% of the total decay rate. If the basis is changed to $qcb$ with mixing neglected, this decay is not allowed. The small mixing that occurs in this basis (or the large mixing in the $cbq$ basis) is what allows there to be any rate in this channel. The net effect is that this exclusive decay is suppressed by a factor of more than 17 (the suppression factor is more than 30 for the corresponding decays of the $\Omega_{bc}$), and the total decay rate of the $\Xi_{bc}$ is suppressed by a factor of 2. For the $\Omega_{bc}$, the suppression of the total rate is somewhat larger than this.

Among the decays of the $\Xi_{bb}$ and $\Omega_{bb}$, hyperfine mixing modifies the exclusive decay rates but leaves the total decay rate of each of these states essentially unchanged. For the $\Xi_{bc}$ and $\Omega_{bc}$, this mixing significantly alters the total decay rates. However, the total decay rate of the $\Xi_{bc}$ and $\Xi_{bc}^\prime$ taken together is essentially unchanged by mixing. The total decay rate from these two parent baryons into the $\Xi_{cc}$ final state is largely unchanged (2.46, 2.44 and 2.45 ($\times 10^{10} s^{-1}$) for the $cbq$ unmixed, $qcb$ unmixed and $cbq$ mixed representations of the $\Xi_{bc}$ states, respectively). The same is true for the total decay rate into the $\Xi_{cc}^*$, with rates of 14.90, 15.10 and 15.08 $\times 10^{10} s^{-1}$ from the three representations of the $\Xi_{bc}$ states, respectively, and this pattern is repeated for the decays of the $\Omega_{bc}$ states.

As was suggested in \cite{pr}, and in the work of Bernotas and Simonis \cite{bernotas}, we have found that hyperfine mixing in the wave functions of baryons containing two heavy quarks have significant effects on their semileptonic decay rates. Similar effects can be expected in the strong and electromagnetic transitions of these states, although the effects may not be quite as dramatic as shown in this manuscript. 

\acknowledgments

This work is supported by the Department of Energy, Office of Nuclear Physics, under contract no. DE-AC02-06CH11357 (MP). WR is grateful to the Department of Physics, the College of Arts and Sciences and the Office of Research at Florida State University for partial support. The authors are grateful to J. Goity for useful discussions.

\end{document}